\documentclass[useAMS,usenatbib,usegraphicx]{mn2e}

\def\rsun{R$_{\sun}$}
\def\aap{A\&A}
\def\apjl{ApJ}
\def\apj{ApJ}
\def\apjs{ApJS}
\def\aj{AJ}
\def\mnras{MNRAS}
\def\nat{Nature}
\def\pasp{PASP}

\title[A Disk Around PG1541+651]
{The Discovery of a Debris Disk Around the DAV White Dwarf PG 1541+651}

\author[M. Kilic et al.]
       {Mukremin Kilic$^{1,4}$,
       Adam J. Patterson$^1$,
       Sara Barber$^1$,
       S. K. Leggett$^{2,4}$,P. Dufour$^3$\\
       $^1$Homer L. Dodge Department of Physics and Astronomy, University of Oklahoma,
       440 W. Brooks St., Norman, OK, 73019, USA\\
       $^2$Gemini Observatory, 670 N. A'ohoku Place, Hilo, HI 96720, USA\\
       $^3$D\'epartement de Physique, Universit\'e de Montr\'eal, C.P. 6128, Succursale Centre-Ville,
       Montr\'eal, Qu\'ebec H3C 3J7, Canada\\
       $^4$Visiting Astronomer at the Infrared Telescope Facility
}

\begin{document}

\maketitle

\begin{abstract}

To search for circumstellar disks around evolved stars, we targeted roughly 100 DA white
dwarfs from the Palomar Green survey with the Peters Automated Infrared Imaging Telescope (PAIRITEL).
Here we report the discovery of a debris disk around one of these targets, the pulsating white
dwarf PG 1541+651 (KX Draconis, hereafter PG1541).
We detect a significant flux excess around PG1541 in the $K$-band. Follow-up
near-infrared spectroscopic observations obtained at the NASA Infrared Telescope Facility (IRTF)
and photometric observations with the warm {\em Spitzer Space Telescope}
confirm the presence of a warm debris disk within 0.13-0.36 \rsun\ (11-32$\times$ the stellar radius)
at an inclination angle of $60^{\circ}$. At $T_{\rm eff}$ = 11880 K, PG1541 is almost
a twin of the DAV white dwarf G29-38, which also hosts a debris disk. All previously known dusty
white dwarfs are of the DAZ/DBZ spectral type due to accretion of metals from the disk. High-resolution
optical spectroscopy is needed to search for metal absorption lines in PG1541 and to constrain the
accretion rate from the disk. PG1541 is only 55 pc away from the Sun and the discovery of
its disk in our survey demonstrates that our knowledge of the nearby dusty white dwarf population
is far from complete.

\end{abstract}


\section{INTRODUCTION}

Eighteen years after the discovery of the first dusty white dwarf \citep[G29-28,][]{zuckerman87}, \citet{kilic05}
and \citet{becklin05} identified the second white dwarf with a circumstellar debris disk, GD 362. The latter was
the most metal-rich white dwarf known at the time \citep{gianninas04}. The discovery of circumstellar disks around
these two DAZ stars led to a paradigm shift in our understanding of the source of the metals in white dwarf photospheres
\citep{kilic07} and also fueled further searches for debris disks around metal-rich white dwarfs. The discovery of
the third \citep{kilic06} and fourth \citep{vonhippel07} dusty white dwarfs followed shortly. Thanks to the
{\em Spitzer Space Telescope}, we now know of $>$20 white dwarfs with circumstellar disks
\citep[see][and references therein]{farihi10,farihi11,dufour10,debes11,melis11,xu11,girven11,steele11}. A detailed photospheric abundance analysis
of these white dwarfs show that the accreted metals are likely to originate from tidal disruption of minor planets
with compositions similar to that of Bulk Earth \citep{zuckerman07,klein10,klein11,dufour10}.

All of the previously known dusty white dwarfs show metal-absorption features in their optical spectra (DAZ or
DBZ spectral type), and they were specifically targeted (except G29-38) for infrared searches because they are metal-rich.
The best way to identify new dusty white dwarfs would be to observe more DAZ/DBZ white dwarfs in the infrared. 
However, the classification of a DA as a metal-rich star requires the detection of weak metal lines with equivalent
widths of $\sim$ 50 mA. This can only be achieved with high resolution spectroscopy at 10m class telescopes
\citep{zuckerman03,koester05}. A cheaper way to find more dusty white dwarfs is to perform a near-infrared search
around a large sample of white dwarfs without prior knowledge of metal absorption features in the targeted stars.
With high-quality optical and near-infrared photometry, a warm debris disk signature can be recognized in the
$K$-band \citep{kilic06}. For example, a $K$-band only
search would detect the disks around 70\% of the known dusty white dwarfs.

To identify new dusty white dwarfs and to constrain the frequency of dusty white dwarfs among the DA population,
we performed a near-infrared $JHK$ survey of $\approx$100 DA white dwarfs from
the Palomar Green (PG) survey using the robotic 1.3m telescope PAIRITEL \citep{bloom06}.
\citet{liebert05} performed a detailed model atmosphere analysis of the DA white dwarfs in the PG survey
using optical spectroscopy and provide temperature, surface gravity, mass, and age estimates. Their
spectroscopy does not have enough resolution to detect the metal-rich DAZs. We select
$\approx$100 apparently single white dwarfs with $T_{\rm eff}=$ 9000-22,000 K, where we are most
efficient in finding the disks using near-infrared data.

Here we report the discovery of a circumstellar debris disk around one of our targets,
the DA Variable (DAV) white dwarf PG1541. The rest of our sample will be presented in an upcoming publication.
In Section 2 we describe our infrared photometric and spectroscopic observations.
In Section 3 we present the spectral energy distribution of PG1541 and constrain
the physical parameters of the circumstellar disk. We discuss the population of dusty
white dwarfs in the PG survey in Section 4.

\section{OBSERVATIONS}

\subsection{Near-Infrared Photometry}

PG1541 is a 510 Myr old ZZ Ceti variable white dwarf with $T_{\rm eff}=11880$ K, $\log{g}=8.2$, $M=0.73M_{\odot}$, $R=0.011 R_{\odot}$ and
$d= 55$ pc \citep{gianninas11}. It was originally an A type star with an initial mass of 3 $M_{\odot}$
\citep{kalirai08,williams09}. It has an apparent magnitude of $g=15.6$ in the Sloan Digital Sky Survey (SDSS). It is detected
in the Two Micron All Sky Survey \citep[2MASS,][]{skrutskie06} with $J = 15.604 \pm 0.062, H= 15.912 \pm 0.171$, and
$K_{\rm s}= 15.429 \pm 0.175$ mag. Due to inaccurate $H$- and $K_{\rm s}$-band photometry,
it is impossible to search for an infrared excess around this object using the 2MASS
data alone. We obtained simultaneous $JHK_{\rm s}$ imaging of PG1541 using the PAIRITEL on UT 2010 February 15. 
PAIRITEL is the old 2MASS telescope and it uses the same camera and the filter set as 2MASS. We use the
PAIRITEL Pipeline version 3.3 processed images and standard IRAF DAOPHOT routines to perform aperture photometry on every
2MASS source detected in the images. We use the 2MASS stars to calibrate the photometry for PG1541. We measure
$J= 15.625 \pm 0.013, H= 15.590 \pm 0.025$, and $K_{\rm s}= 15.450 \pm 0.048$ mag in the 2MASS magnitude system. Our $J$-
and $K_{\rm s}$-band photometry is consistent with the 2MASS measurements within the errors. Thanks to the smaller errors in
the PAIRITEL photometry, a significant flux excess is revealed in the $K_{\rm s}$-band.

\begin{figure}
\begin{center}
\includegraphics[width=2.5in]{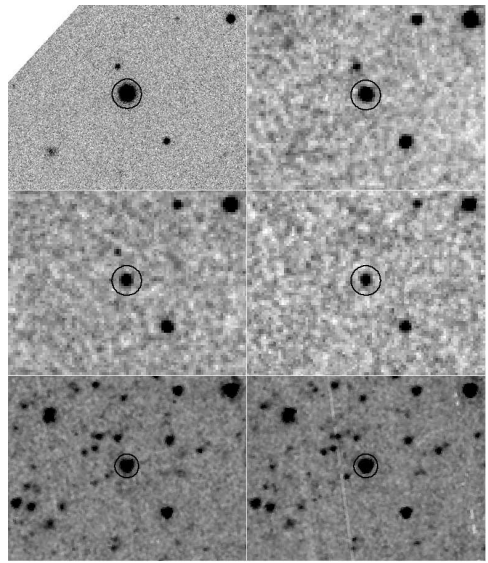}
\end{center}
\caption{The SDSS $g$-band and PAIRITEL $J$ (top panel), $H$ and $K_{\rm s}$ (middle panel), and IRAC 3.6 and 4.5$\mu$m (bottom panel)
images of PG1541.}
\end{figure}

\begin{table}
\centering
\caption{Optical and Infrared Photometry of PG1541}
\begin{tabular}{lcr}
\hline
Filter & $\lambda_{\rm eff}(\mu$m) & $F_{\nu}(\mu$Jy) \\
\hline
$FUV$   & 0.15 &  574.4 $\pm$ 25.3 \\
$NUV$   & 0.23 & 1330.3 $\pm$ 25.7 \\
$u$     & 0.36 & 1567.4 $\pm$ 29.4 \\
$g$     & 0.47 & 2186.9 $\pm$ 41.5 \\
$r$     & 0.62 & 1971.6 $\pm$ 32.4 \\
$i$     & 0.75 & 1677.6 $\pm$ 22.4 \\
$z$     & 0.89 & 1273.1 $\pm$ 16.9 \\
$J$     & 1.24 &  896.4 $\pm$ 19.0 \\
$H$     & 1.66 &  594.7 $\pm$ 18.0 \\
$K_{\rm s}$ & 2.16 &  440.5 $\pm$ 21.2 \\
$IRAC1$ & 3.56 &  435.4 $\pm$ 13.5 \\
$IRAC2$ & 4.51 &  483.9 $\pm$ 15.0 \\
\hline
\end{tabular}
\end{table}

\subsection{Near-Infrared Spectroscopy}

We obtained low resolution near infrared spectra of PG 1541 on 2011 April 13 using the 3m NASA Infrared
Telescope Facility (IRTF) equipped with the 0.8--5.4 Micron Medium-Resolution Spectrograph and Imager
\citep[SpeX,][]{rayner03}. The observing setup and procedures are similar to those of \citet{kilic08}.
We use a $0.5\arcsec$ slit to obtain a resolving power of
90--210 over the 0.7--2.5 $\mu$m range. The observations are taken in two different
positions on the slit separated by 10$\arcsec$. The total exposure time for PG1541 is 32 min.
We use internal calibration lamps (a 0.1W incandescent lamp and an Argon lamp)
for flat-fielding and wavelength calibration, respectively. To correct for telluric features
and flux calibrate the spectra, we use the observations of the nearby bright A0V star HD 143187
at an airmass similar to the PG1541 observations. We use the IDL-based package Spextool version 3.4
\citep{cushing04} to reduce the data.

\subsection{{\em Spitzer} Photometry}

We used the warm {\em Spitzer} equipped with the InfraRed Array Camera \citep[IRAC,][]{fazio04} to
obtain infrared photometry of PG1541 on UT 2011 January 6 as part of
the program 70023. We obtained 3.6 and 4.5$\mu$m images with integration times of 30 seconds for nine dither positions.
Figure 1 shows the optical and infrared images of the field around PG 1541. These images reveal
no contaminating sources around PG1541.
We use the IDL astrolib packages to perform aperture photometry on the individual
corrected Basic Calibrated Data frames from the S18.18.0 pipeline reduction.
We get similar results using 2, 3, and 5 pixel apertures, but we quote the results using the smallest aperture since
it has the smallest errors.

Following the IRAC calibration procedures, we correct for the location of the source in
the array
before averaging the fluxes of each of the dithered frames at each wavelength.
We also correct the Channel 1 (3.6$\mu$m) photometry for the pixel-phase-dependence.
We estimate the photometric error bars from the observed scatter in the nine images
corresponding to the dither positions. We add the 3\% absolute calibration error in quadrature \citep{reach05a}.
\citet{reach09} demonstrate that the color corrections for dusty white dwarfs like G29-38 are small (0.4-0.5\%) for
channels 1 and 2. We ignore these corrections for PG1541.
We measure F$_\nu = 435.4 \pm 13.5 \mu$Jy in Channel 1 and  $483.9 \pm 15.0 \mu$Jy in Channel 2.
Table 1 presents the Galaxy Evolution Explorer \citep[GALEX,][]{morrissey07} ultraviolet, SDSS optical,
and PAIRITEL/{\em Spitzer} infrared photometry of PG1541. We use the corrections given in \citet{eisenstein06}
to convert the SDSS photometry to the AB magnitude system.

\section{RESULTS}

Figure 2 presents the flux calibrated spectrum of PG1541 along with the optical photometry from the
SDSS and the PAIRITEL near-infrared photometry. The observed spectrum is shown as a jagged line, and
the expected near-infrared photospheric flux from the star \citep{bergeron95} is shown as a solid line.
The model spectrum for PG1541 is calculated by one of us (P.D.) using the parameters given in \citet{gianninas11}.
To match the resolution of the IRTF spectrum, the model white dwarf spectrum is shown at 100 \AA\ resolution.
The dotted line shows the telluric spectrum obtained from the reference A0V star observations.
The telluric correction for PG1541 is not perfect in the regions around 1.4 and 1.9 $\mu$m.

\begin{figure}
\includegraphics[width=3.5in]{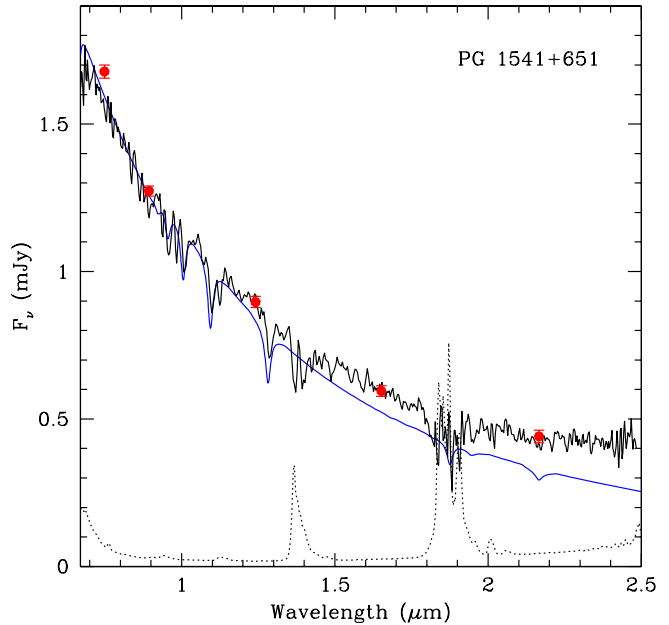}
\caption{Flux calibrated spectrum of PG1541 (jagged line) and the telluric and instrumental spectrum
derived from the reference A0V star (dotted line). The expected flux from the stellar photosphere is shown
as a solid line. SDSS optical photometry and PAIRITEL near-infrared photometry are shown as filled points.}
\end{figure}

The IRTF spectrum matches the observed photometry in the 0.7-2.5$\mu$m region relatively well and it confirms
the slight $H$-band excess and more significant $K_{\rm s}$-band excess in the PAIRITEL photometry. The shape of the
infrared excess is similar to the excess seen in the other dusty white dwarfs; there are no obvious
emission or absorption features, and it is rising toward longer wavelengths. 

Figure 3 shows the spectral energy distribution of PG1541 in the 0.1-5$\mu$m range, including the
GALEX, SDSS, PAIRITEL, and {\em Spitzer} photometry. PG1541 is relatively nearby and the line of sight
reddening is only $E(B-V)=0.03$ \citep{schlegel98}; its spectral energy distribution does not show any
evidence of reddening.

PG1541 is brighter at 4.5$\mu$m than it is at 2$\mu$m.
Since the stellar contribution falls like a power law in this wavelength range, the contribution
from the cool component is still rising at 5$\mu$m. This contribution is almost identical to the excess flux
seen around G29-38 in the same wavelength region \citep[see Figure 3 in][]{reach09}. A brown
dwarf companion cannot explain the observed photometry in the 2-5 $\mu$m range \citep{leggett10}.
In addition, spectral features from such a companion would be visible in our IRTF spectrum.
Hence, the infrared excess around PG1541 can only be explained by a circumstellar disk.

\begin{figure}
\includegraphics[width=3.5in]{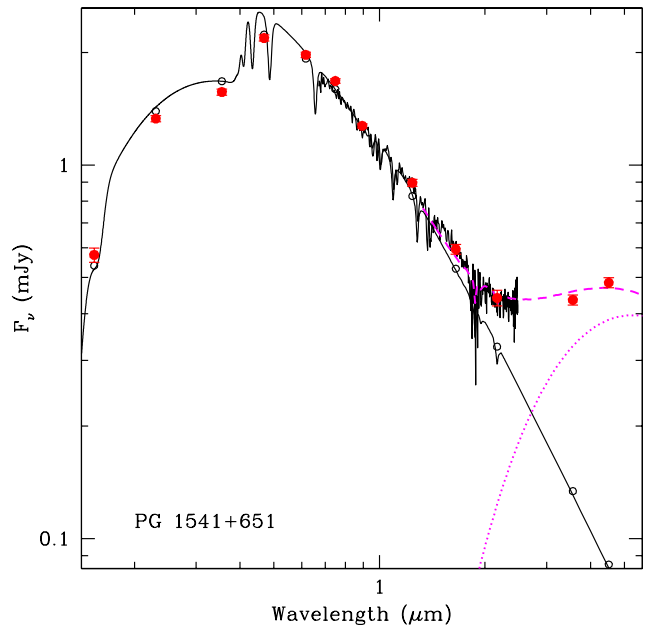}
\caption{Spectral energy distribution of PG1541 from ultraviolet to infrared wavelengths. The GALEX, SDSS,
PAIRITEL, and IRAC photometry as well as our IRTF spectrum are shown. The solid line represents
the contribution from the stellar photosphere and the open circles represent the average over the filter bandpasses.
The dotted line shows the contribution from the debris disk.}
\end{figure}

We fit the infrared excess around PG1514 using the optically thick flat-disk models of \citet{jura03}. Using
the effective temperature, radius, and distance estimates for the white dwarf, we created a grid of disk models with
inner temperatures 1000-1800 K, outer temperatures 300-1300 K (in steps of 100 K and with the condition that
$T_{\rm in} > T_{\rm out}$) and inclination angles $0^{\circ}-90^{\circ}$ (in steps of 10$^{\circ}$). The model that best
fits the infrared photometry has $T_{\rm in}$ = 1300 K, $T_{\rm out}$ = 600 K, and an inclination angle of $i=60^{\circ}$.
To estimate the uncertainties for these parameters, we perfom a Monte Carlo analysis where we replace
the observed photometry $f$ with $f + g\delta f$, where $\delta f$ is the error in flux and $g$ is a Gaussian
deviate with zero mean and unit variance. For each of the 10,000 sets of modified photometry, we repeat
our analysis and derive new best-fit parameters for the disk. We adopt the interquartile range as the uncertainty.
This analysis yields $T_{\rm in} = 1300 \pm 100$ K, $T_{\rm out} = 600 \pm 100$ K, and $i=60 \pm 10$ degrees.
The disk is within 0.13-0.36 \rsun\ of the white dwarf; this is within the Roche radius for tidal disruption of
an asteroid \citep{vonhippel07}. These parameters are similar to the parameters for the other
dusty white dwarfs studied with the same flat-disk model, including G29-38 \citep{jura03}.

The total mass of the disk cannot be constrained for an opaque ring. Based on the choice of optically thin
or thick models, the disk around G29-38 holds $10^{19}-10^{24}$ g of material \citep{jura03,reach05b}. A similar
amount of material may exist in the disk around PG1541.

\section{DISCUSSION}

Along with GD 133, WD 1150$-$153, and G29-38, PG1541 is the fourth DAV known to host a circumstellar debris disk.
Every DA is expected to go through the ZZ Ceti instability strip during its evolution \citep[see
the discussion in][]{gianninas11}. Hence, we expect that some of the DAZs with disks will also be DAVs;
there is nothing special about finding 4 dusty DAV white dwarfs. However,
optical and infrared observations of the pulsations in the first known dusty white dwarf, G29-38, were important for
identifying the source of the infrared excess as a debris disk rather than a brown dwarf companion \citep{graham90}.

The main pulsation mode in PG1541 has an amplitude of 4.5\% in the optical with a period of 689 s \citep{vauclair00}.
\citet{reach09} detect 4\% fluctuations in G29-38 at 3.6$\mu$m, whereas the main pulsation mode in the optical has
an amplitude larger than 20\% \citep{mcgraw75}. The infrared variations are likely due to the dust grains going
through temperature fluctuations with the stellar pulsations. Assuming that the disk around PG1541 is similar to
G29-38's disk, we do not expect to see any significant ($>$1\%) variations in our IRAC data for PG1541. Hence our disk
model is not likely to be effected by stellar variability.

Previous {\em Spitzer} surveys find a frequency of 1-3\% for dusty white dwarfs \citep{mullally07,farihi09,kilic09}.
Out of the 348 DAs analyzed by \citet{liebert05}, there are now six PG white dwarfs known to host debris disks:
PG 0843+517, 1015+161, 1116+026, 1457$-$086, 1541+651, and 2326+049 (G29-38). Therefore, the frequency of debris disks
in the PG sample is at least 1.7\%. However, not all of the DA white dwarfs in the PG survey have been observed in the infrared,
and a more accurate analysis will have to wait until our entire sample is presented in an upcoming publication.

The ongoing Wide-field Infrared Survey Explorer \citep[WISE,][]{wright10} mission should greatly increase
the sample of known dusty white dwarfs in the solar neighborhood. The WISE InfraRed Excesses around Degenerates (WIRED) survey
has already identified a new dusty white dwarf at 55 pc, GALEX1931 \citep{vennes10}, and
it is predicted to detect disks around G29-38-like objects within
$\sim$100 pc of the Sun \citep{debes11}. 
However, due to its relatively large beam size ($\geq6.1\arcsec$), background
contamination may be a problem for WISE observations of targets in crowded fields \citep{melis11}.
PG1541 is bright enough to be detected in the WISE 3.4 and 4.6 $\mu$m bands, and
possibly in longer wavelength data if it displays a silicate emission feature.

\section{CONCLUSIONS}

We have detected an infrared excess at 2-5 $\mu$m around the 510 Myr old (cooling age) pulsating
white dwarf PG 1541+651 and determined it to be
a narrow dust ring within the tidal disruption radius of the white dwarf. Near-infrared spectroscopic data show
a significant contribution from a featureless spectrum rising to the red, confirming our dust disk interpretation.
PG1541 is most likely a DAZ, however high-resolution optical spectroscopy is required to identify weak metal
absorption features. 

We detected the disk around PG1541 in our survey of $\approx$100 DA white dwarfs from the PG survey. Our remaining targets
will be presented in a future publication, which will also address the frequency of disks in the PG survey.

\section*{Acknowledgements}
P.D. is a CRAQ postdoctoral fellow.
The PAIRITEL is operated by the Smithsonian Astrophysical Observatory (SAO)
and was made possible by a grant from the Harvard University Milton Fund, the camera loan from the University of Virginia,
and the continued support of the SAO and UC Berkeley. The IRTF is operated by the University of Hawaii under Cooperative
Agreement no. NNX-08AE38A with the National Aeronautics and Space Administration, Science Mission Directorate, Planetary
Astronomy Program.

\end{document}